# The Nanotechnology R(evolution)[1]

Charles Tahan
*Cavendish Laboratory, University of Cambridge,
JJ Thomson Ave, Cambridge, CB3 0HE, UK*
charlie@tahan.com
(2006)

**Abstract**

Nanotechnology as a social concept and investment focal point has drawn much attention. Here we consider the place of nanotechnology in the second great technological revolution of mankind that began some 200 years ago. The so-called nanotechnology revolution represents both a continuation of prior science and technology trends and a re-awakening to the benefits of significant investment in fundamental research. We consider the role the military might play in the development of nanotechnology innovations, nanotechnology's context in the history of technology, and the global competition to lead the next technological revolution.

**Table of Contents**

## 1. Introduction

Roughly seven thousand years ago, humans began to leave their nomadic ways and form civilizations around the irrigation and cultivation of land. As a result, human society and community transformed radically. The creation of government and bureaucracy, of social classes, written language, the rule of law, the notion of the individual, standing armies, and much more, all emanated from this technological change. Dubbed the "irrigation society" by renowned management thinker Peter Drucker (Drucker, 1965), this first great technological revolution of man lasted over two thousand years.

Nowadays, the word revolution is used rather freely. From the "internet revolution" to the "digital music revolution" to the "nanotechnology revolution," at what scale does an innovation become more than an innovation? In her classic work *Technological Revolutions and Financial Capital* (2002) Carlota Perez defines five technological revolutions since the end of the eighteenth century. As new technologies emerge and disseminate, they tend to follow similar economic investment cycles which Perez calls "techno-economic paradigms." However, the key realization – and what Drucker was suggesting in his 1965 presidential speech to the Society of the History of Technology – is that we are living in a second great technological revolution. Beginning with the Industrial Revolution in Britain (around 1750) a man could expect to die in a world very much different from the one into which he was born.

---

[1] Preprint of chapter to appear in *Nanoethics: Examining the Societal Impact of Nanotechnology*, Fritz Allhoff, Patrick Lin, James Moor, and John Weckert, eds., (2007).

A great revolution implies an increasing rate of innovation, and not just technological innovation, but also organizational and political, in many different fields. There is little doubt that the rate of innovation accelerated after 1750 (Cross and Szostak, 1995) and is still increasing. Although it is difficult to assess the evolution and longevity of such a revolution from within it, we can easily believe that as more people and more wealth come into the enterprise of innovation, more of Perez's techno-economic paradigms will occur with increasing frequency. Nanotechnology may be one of them.

Technological revolution is always accompanied by political and social change. The two go together. The bigger the technological change, the more society must adapt to accommodate this objective reality. For example, as farmers began to accumulate wealth (that is, food) for communities, an army became necessary to protect it. Just as there are larger and smaller technological revolutions, there are larger and smaller societal changes to go along with them. In Table 1 we have tried to form a hierarchy of revolutions. Inevitably, revolutions are coarse-grained, representing a sum of individual innovations that erupt seemingly randomly. We define four categories of revolution: great revolutions, of which there have only been two, with the present one just beginning; major revolutions or the techno-economic paradigms of Perez, which usually last some 50 years; minor revolutions, which are finer still and are key building blocks of the major revolutions; and micro revolutions, which are largely new investment opportunities in technology that come about within the larger revolutions but also follow a cyclical pattern of investment and saturation.

Soon after World War II, governments worldwide and particularly in the US realized that significant investment in the natural sciences could drastically affect the power and wealth of nations. Integrated electronics, the Internet, even lasers can be lumped into the age of information and telecommunications as a direct result of this investment. Much of what is today called nanotechnology naturally follows from these lines of technical pursuit and scientific inquiry. Together with the theoretical understanding of quantum physics and electrodynamics that has developed over the last century, this continuation of research has led us within reach of tremendous rewards from the manipulation and control of matter at the nanoscale. Whereas these pursuits were somewhat ignored in recent decades – significant hype and investment focusing instead on telecommunications, software and networking, and biotechnology – there is a growing realization that it is time to start pushing materials science and fundamental research again. Very recent trends in the global energy crisis and the so-called green revolution only add to this notion.

If anything, nanotechnology has become a marketing term to encompass and drive this belief that more funding is needed in the physical sciences to maintain economic, scientific, and military advantage over international competition. As evidence, roughly one-third of the budget for the National Nanotechnology Initiative (NNI) this year will go to the National Science Foundation (NSF) (Roco, 2004), which primarily supports unfettered basic research. Still, understanding how government and the military drive technological development and how nanotechnology as it stands today (and may exist in the future) may relate to prior revolutions in history has great value. Responsible encouragement of the great technological revolution in which we find ourselves is vital to human civilization.

**Table 1: Hierarchy of technological revolutions**

*Great Revolutions* (there have only been two that we know about)
    $1^{st}$: Irrigation society, began – ended: approximately 5000 BC - 3000 BC
    $2^{nd}$: Began with the Industrial Revolution in Britain in the $18^{th}$ century
*Major Revolutions* after 1750 (start date)
    the industrial revolution (1771)
    the age of steam and railways (1829)
    the age of steel, electricity and heavy engineering (1875)
    the age of oil, the automobile and mass production (1908)
    the age of information and telecommunications (1971)
    *the age of bio-engineering (1980)?*
    *the second industrial revolution (1991)?*
    *the age of machine-phase nanotechnology (2030-50)?*
*Minor Revolutions* (some examples)
    personal computing
    mobile phones
    global networking
    *nanoparticle revolution?*
*Micro Revolutions* (some examples)
    digital music revolution
    HD TV revolution
    *nanoparticle revolution?*

## 2. Defining nanotechnology

Nanotechnology is a social construction. The word nanotechnology did not emerge as a distinct area of science, but rather was introduced externally and defined by its usage in the greater societal dialogue. A primary consequence of this very public defining of the term nanotechnology is it's present bipolar nature. We take a pragmatic definition of nanotechnology that combines both sides: the *reality* of the word as it is used today primarily by governments, corporations, and scientists as well as the *vision* of what the field might become (Tahan, 2007). The reality of nanotechnology – is defined mostly by government funding managers and agencies – largely encompasses ongoing research in materials science and solid-state physics. Examples include nanoparticles and quantum dots, "nano-enabled" surface coatings, transistor features that are less than 10 nm scale, giant- and colossal-magneto resistance (as in hard drives), spintronics, photonic band-gap structures, and more. The definition usually takes a variant like this one from the Royal Society of the UK (Royal Society, 2004): "nanoscience is the study of phenomena and manipulation of materials at atomic, molecular, and macromolecular scales, where properties differ significantly from those at a larger scale." Distinct from this is the more science fiction *vision* of nanotechnology popularized by Eric Drexler (Drexler, 1986) and in books like Neal Stephenson's *The Diamond Age* (Stephenson, 1995), that of atom-by-atom construction of matter and nanoscale (invisible) machines and robots, also referred to as "machine-phase nanotechnology."

The *reality* definition of nanotechnology is a synonym for fundamental materials and matter research, including quantum phenomena at small length scales. The origins are clear. For the past 40 years, since the invention of the semiconductor transistor, the economic apparatus built around Moore's Law has been driving material features ever smaller. Concurrently, our understanding of the basic quantum physics that governs the behavior of interacting particles (of matter and light) has been solidifying. More recently, measurement and fabrication techniques have reached a point where we can start thinking seriously about exploiting some of these novel properties that appear in the small length regime. As this level of control gets closer technologically, with clear opportunities in sight, the funding of these endeavors becomes more worthwhile. If nanotechnology can act as an umbrella term to drive interest and funding, so be it. The head of the NNI publicly espouses this viewpoint (Roco, 2004).

We can further separate the larger field nanotechnology, in terms of the reality and vision accompanying this emerging technology, from the recent interest in nanoparticle and quantum dot technologies. In many ways, certainly as far as environmental and human toxicity (Colvin, 2003) are concerned, nanotechnology can be defined much more narrowly than the above (as we have argued before (Tahan, 2007)):

> Nanotechnology, at present, is nanoparticles and nanomaterials that contain nanoparticles. Nanoparticles are defined as objects or devices with at least two dimensions in the nanoscale regime (typically tens of nanometers or less) that exhibit new properties, physical, chemical, or biological, or change the properties of a bulk material, due to their size. Nanotechnology of the future will include atom-by-atom or molecule-by-molecule built active devices.

Much of the excitement surrounding nanotechnology comes from the promise of newly gained nanoparticle synthesis techniques and a realization of their potential in many different areas. Two prominent examples are bio-markers for cancer detection (and destruction) and quantum dots in solar-energy conversion devices (Scientific American, 2002). Nanoparticle and nanoparticle-composite technology may end up solely a minor or micro revolution separate from the broader nanotechnology field

### 3. Nanotechnology's place in an age of ages

*It is not only the speed of technological change that creates a "revolution," it is its scope as well. Above all, today, as seven thousand years ago, technological developments from a great many areas are growing together to create a new human environment.* (Drucker, 1965)

Drucker's words remain true. Nanotechnology as presently (loosely) defined will likely have several acts to play in the coming century. Certainly the utilization of nanoparticle technology has immediate promise. Much of the rest of nanotechnology in the near term can more accurately be placed within the information or biotechnology revolutions. In either case, we can find patterns. The five major revolutions that Perez outlined all follow a similar pattern. The first stage is the installation period, which has an eruption phase, when a new innovation is introduced and spreads in conflict with old products and technologies. The second is the frenzy phase, when financial capital drives the build-up of new technologies but develops tensions within the system. A turning point occurs, usually with a recession that follows the collapse of a financial bubble, and regulatory changes are made to facilitate and shape the period of development. Then follows a period of deployment, which initially has a synergy phase, when conditions are all favorable for the full flourishing of the new technology, and then the maturity phase, when signs of dwindling investment opportunities and stagnating markets appear (Perez, 2002).

Obviously there is much fluctuation in this model. Since nanotechnology as labeled takes on so many meanings, we must separate the key components. First, there is the nanoparticle/quantum dot component, and we will call this the nanoparticle revolution. Second, there is a continuation of technologies resulting in nanoscale techniques for manufacture that are being widely adopted by big industry (GE, Dupont, Intel). We can hesitantly call this a second industrial revolution (ground up technology?). Finally, in the far distance, there is the machine-phase nanotechnology revolution, completely imaginary at this stage. Only this stage of development (promising essentially free goods) holds the potential for drastic social and political upheaval.

At the current state of development, nanotechnology simply does not represent a paradigm shift in scientists' thinking. Nanoscale investigation is an evolutionary outgrowth of a new capability to measure and fabricate at that scale. Nanotechnology must be seen in the greater trend of innovation, which, like the irrigation revolution, will likely continue well into the next millennium.

## 4. The military and technological development

The military has long been an instigator and shaper of technological innovation. Often the high-cost buyer or buyer of last resort, the military can act both to encourage a fledgling technology and to prolong a dying one. The US Department of Defense (DOD) has clearly taken an interest in nanotechnology and accounts for roughly 28% of all federal funding in the loosely defined field in FY2005 (Roco, 2004). One prominent example is the Institute for Soldier Nanotechnologies (Talbot, 2002) at the Massachusetts Institute of Technology (http://web.mit.edu/ISN/). Stronger and lighter materials and more explosive bombs (super thermites) are but two examples of nanotechnology's impact on future warfare.

It is widely assumed that military-born technologies spill over into civilian use for beneficial purposes. However, the military has specific objectives in its approach to technology, which might be very different from those of society at large. Historian David Noble lays out three such objectives in his treatise on military and technology (Noble, 1987) which are worth considering again in the context of nanotechnology. These are 1) performance (emphasis placed on meeting military objectives and what follows necessarily from them), 2) command (management techniques with decision-making coming solely from the top), and 3) modern methods (a fetish for machinery that won't talk back). Noble argues that it is a misconception that the military acts only as an external input of technology. Instead, the military shapes the progress and nature of a technology or set of technologies throughout their lifetime in many cases.

One clear example of this dates back to the beginning of the US as a nation where the military's quest for interchangeable gun parts helped spur mechanization and the industrial revolution in the States. Uniformity was imposed by the military contract system. "The benefits of the system, clear to the military, were not so clear to many manufacturers, given the high costs, uncertainties, and inescapable industrial conflict it engendered" (Noble, 1987). A similar example is that of *numerical control*, pioneered by the Air Force in the 1980s. Numerical control envisioned extremely precise machining based on computer and mathematical specification and extreme shortening of the chain of command from aircraft part specification to manufacture. Industry generally was not enthusiastic as the systems were very complex and not as flexible as other less-demanding, though adequate, metal working techniques. Since the military provided such a large and stable base of funding, however, industry followed the numerical control path (Noble, 1987). Industry paid the price as foreign competition became more nimble. The loss of promising alternative technologies, excessive consolidation in the metalworking industry, and slow innovation all resulted from the military's involvement.

An excellent counter-example to this phenomenon is Intel. Although the military was the initial buyer of Intel's first few-transistor circuits, its preferences did not shape Intel's future. Intel would not have survived in the rapidly changing consumer environment had its engineers been unable to make decisions. In fact, Intel has thrived on a very long chain of command. In other words, engineers very near the technology (but at the bottom of the corporate hierarchy) are entrusted with a large amount of discretion to make decisions related to technology undeniably vital to the company's future. The rate of growth in the private sector made this possible, although it is important to note that some semiconductor fabs continued at a reduced level by specializing their wares for military needs (think Fairchild Semiconductor, a founding company of silicon valley which has since largely left the commercial consumer electronics arena for mostly military and advanced technology contracting).

The long-term trend is an increasing shift of federal research dollars into the mission-driven agencies and away from discipline-driven research, such as in NSF, the Department of Energy's (DOE) Office of Science, and at NIST (National Institute of Standards and Technology). Apart from the National Institutes of Health (NIH), non-defense federal R&D is about the same in 2004 dollars as it was in 1980 (Duke and Dill, 2004). Basic, unfettered physical research in the US is declining, except where it goes through the DOD mission agencies and the NNI.

In general, the military funding agencies are much stricter about how their grant money is used as compared to the NSF. Since all or most of the research money comes through the military, and they are the ones asking hard questions and threatening to pull funding, scientists at universities feel strong pressure to follow the dictated "roadmaps" instead of pursuing new physics as it is identified. Because of this, new and perhaps useful phenomena at the nanoscale – which may lay the groundwork for the next revolution 50 years hence – may be missed in this country.

## 5. Lessons from the past

A recent commercial by General Electric featured a "professor of nanotechnology" and a super model falling in love: "the perfect combination of brains and beauty" (GE, 2005) "Nanotechnologist" has become the new computer scientist, driver of the next great wealth generator. While easy to dismiss, it is important to remember that but for the abnormal obsession of a couple dozen people with the properties of semiconductors, the US would not have led the personal computing, networking, and internet revolutions of the last half of the 20$^{th}$ Century. A case study is Great Britain, who irreversibly fell behind Germany and the US because it faltered in its investment in new technologies during the age of steel, electricity, and heavy engineering (1875-1920) (Duke and Dill, 2994). Will history repeat itself in the US?

Before World War II, US universities were a joke internationally. Due to the demonstrable success of radar and the atomic bomb, the US quickly realized that science played a key role in military victories and national power, so a large-scale investment in fundamental research began. The GI Bill supplied manpower. Through this and America's survival as a superpower after the war (and a concurrent influx of highly trained European scientists), the US has had the good fortune to lead the last major techno-economic revolution: the age of information and telecommunications, as well as many minor ones. But as other countries catch up to America's core strengths, the US leadership position is tenuous. Indeed, funding in nanotechnology in Japan and Europe is comparable to that of the US at present (Roco, 2004; NNI Triennial Review 2006).

If we want to lay the foundation for the next revolution, it is instructive to go back and try to understand what began in the mid 18$^{th}$ century in England. In fact, the name Industrial Revolution is a misnomer, as innovations took place in many areas such as farm and home, in addition to manufacturing (Cross and Szostak, 1995). No one knows for sure why the Industrial Revolution began where and when it did. There are many hypotheses: Britain's institutional support of technology (world's first patent system, strong private property rights, acceptance of Jewish and other ethnic minorities); urbanization and increased life expectancy; encouragement of an empirical and utilitarian tradition (e.g., Francis Bacon's writings); consolidation of agricultural land by lords with agricultural efficiencies; increased worker migration to cities; movement of work away from the guild system (putting out system); raw material advantages; less regulation. The list goes on. But other European countries like Germany and France, who also had better educational systems, shared many of these advances in whole or in part. The one thing other countries lacked was a transport system even remotely comparable to what England had put in place. "Transport improvements greatly accelerated the processes of regional specialization and urbanization in England. They also led to a dramatic increase in personal travel" (Cross and Szostak, 1995). This encouraged the interaction between innovators with varied backgrounds, expertise, and ideas, which is essential to the innovation process. Regionalization and localization led to mechanization.

In the present day, the US has a mixed infrastructure in idea transportation. Although it has pioneered advancements in collaboration and interaction online, several other countries, such as South Korea or Taiwan have superior broadband networking penetration. Residents of the US have always enjoyed freedom to move about the country, and career success often demands it. The US also enjoys the benefits of scale, with a large number of excellent yet independent universities and a large entrepreneurial culture (exhibited in individuals and in organizations such as top-notch private-equity entities). Presently, first-world countries like Great Britain, Australia, and Japan – although investing heavily in nanotechnology research – are struggling to match the US's highly efficient venture capital ecosystem. However, immigration rules since 9/11 have decreased the influx of talent from around the world, traditionally a key driver of science research. But none of this compares to the great experiment that is ongoing in the very nature of science investment in the US.

Industrial science and technology in the US has undergone a dramatic change in recent years, from "Closed Innovation" to "Open Innovation" (Chesbrough, 2003). In essence, the era of industrial research labs is over (Duke and Dill, 2004). Where significant basic research used to occur in the bowls of Bell Labs or Xerox PARC, industry has now focused more on development of near and more economically justifiable engineering. Extreme examples of this are companies like Intel and Cisco, who "outsource" virtually all their research. They leverage their research budgets by partnering with academia and other companies and start-ups. This is different from and in addition to what's usually called outsourcing – the farming out of actual work or jobs (in this case in research and development) to countries such as China and India. "Under Open Innovation, a company's value chain is no longer fully contained within the company, and ideas, people, and products flow across company boundaries, to and from other companies, universities, and even countries" (Duke and Dill, 2004).

This business trend has left only universities and national labs to fulfill the need for basic research in the US. From 1953 to 1996, the fraction of basic research that was performed in universities and federal labs rose from 33% to 61% (Duke and Dill, 2004). Is this enough to make up the difference? We simply do not yet know how this change will affect US competitiveness in the future. "The growth of biotechnology in America is largely a story of seedling ideas that came from academic scientists in research universities, funded by venture capitalists, and manned by bright graduate and postdoctoral students" (Duke and Dill, 2004). Nanotechnology may prove to be the same story, or not.

There are a number of conditions that allowed the Industrial Revolution to move quickly to the US: fast population growth; natural and artificial protection (via the Atlantic Ocean and tariffs); copying and extending prior work (of the British banking system, corporate, and insurance; manufacturing techniques, etc.); relief from bankruptcy (limited liability); legal monopoly over inventions through patent law – with strict granting of patent applications to ensure that only new and useful ideas were patented; lack of guild monopolies; vast natural resources; receptivity to innovation. What country today has the most of these benefits? As Duke and Dill point out (Duke and Dill 2004), the US must focus on its core strengths: innovative and fast-moving companies, talented people, and strength in basic research. With these concerns, the motivation of the NNI to pump money into fundamental research under the cover of nanotechnology seems like a very good move.

While on the surface the business trend to open innovation seems a good way to speed up business and technology growth, it is unclear what the long term affects on the US will be. If the majority of basic research ends up in Asia, can US corporations seriously believe they will be allowed to "manage" and benefit from these new discoveries indefinitely?

## 6.     Conclusions

That we are living through a great technological revolution with no end in sight is clear. What gets murky is our attempt to sub-classify smaller revolutions within this larger landscape of merging innovations. The reality of nanotechnology as it stands today is the continued evolution of prior trends in information and materials science began after World War II. That we are at a point where the exact synthesis of nanoparticles and other nanotechnologies holds great promise for medicine, energy conversion, etc., has only fueled the belief that a renewed surge of investment is needed to harvest these potential technological breakthroughs. We have neither begun to approach the vision of nano-machines and robots that popularized the term nanotechnology, nor to adequately understand the difficulty in getting there. So the great technological revolution that this may imply lies still in waiting for us to discover.

**Acknowledgements**

The title – "Nanotechnology R(evolution)" – originates with a former student, Michael Markovics, in my class of spring 2005 at the University of Wisconsin-Madison: *Nanotechnology and Society* (Tahan, 2006). The author would like to thank members of the Wisconsin Nanotechnology and Society Initiative for useful conversations and especially Greta Zenner for critical reading and editing. The author is supported by a USA National Science Foundation Math and Physical Sciences Distinguished International Postdoctoral Research Fellowship (Award No. DMR-0502047).